\documentclass[10pt,letterpaper]{article}
\usepackage{opex3}
\newcommand{\mic}{~\mu{\rm m}}
\newcommand{\Dm}{\hat{\mathit{D}}}
\newcommand{\kpp}{k^{(2)}}
\newcommand{\neffs}{N_{\rm eff}}
\newcommand{\ld}{L_{\rm D,1}}
\newcommand{\tin}{T_{\rm in}}
\newcommand{\Iin}{I_{\rm in}}
\begin{document}

\title{Optical Cherenkov radiation by cascaded nonlinear interaction: an efficient source of few-cycle energetic near- to mid-IR pulses}

\author{M. Bache$^1$, O. Bang$^1$, B.B. Zhou$^1$, J. Moses$^2$ and F.W. Wise$^3$}

\address{$^1$DTU Fotonik, Department of Photonics Engineering, Technical University of Denmark, DK-2800 Kgs. Lyngby, Denmark\\ $^2$Optics and Quantum Electronics Group, Massachusetts Institute of
Technology, Cambridge (MA) 02139, USA\\ $^3$Applied and Engineering Physics, Cornell University, Ithaca (NY) 14853, USA}

%
%
%
%

\email{moba@fotonik.dtu.dk} 


\begin{abstract}
When ultrafast noncritical cascaded second-harmonic generation of energetic femtosecond pulses occur in a bulk lithium niobate crystal optical Cherenkov waves are formed in the near- to mid-IR. Numerical simulations show that the few-cycle solitons radiate Cherenkov (dispersive) waves in the $\lambda=2.2-4.5\mic$ range when pumping at $\lambda_1=1.2-1.8\mic$. The exact phase-matching point depends on the soliton wavelength, and we show that a simple longpass filter can separate the Cherenkov waves from the solitons. The Cherenkov waves are born few-cycle with an excellent Gaussian pulse shape, and the conversion efficiency is up to 25\%. Thus, optical Cherenkov waves formed with cascaded nonlinearities could become an efficient source of energetic near- to mid-IR few-cycle pulses.
\end{abstract}



\section{Introduction}

Nonlinear optics dawned when second-harmonic generation (SHG) was demonstrated 50 years ago \cite{franken:1961}. Soon after, it was predicted \cite{Ostrovskii:1967} and experimentally demonstrated \cite{Thomas:1972} that cascaded SHG could lead to a Kerr-like nonlinear action. The cascading occurs when the harmonic generation over a propagation length $L$ is strongly phase-mismatched $|\Delta k L|\gg 1$: only a fraction of the fundamental wave (FW) is up-converted to the second harmonic (SH) after a coherence length $\pi/|\Delta k|$, and after another coherence length it is back-converted to the FW. Due to the different FW and SH phase velocities when $\Delta k\neq 0$, the back-converted FW is phase shifted relative to the unconverted FW. On continued propagation the FW effectively experiences this cascade of up- and down-conversions as a Kerr-like nonlinear refractive index change $\Delta n= n_{\rm casc}^I I$ proportional to the intensity $I$. Since $n_{\rm casc}^I \propto -d_{\rm eff}^2/\Delta k$ the phase-mismatch  $\Delta k=k_2-2k_1$ controls both the magnitude and sign of the cascaded nonlinearity \cite{Ostrovskii:1967}. This has been used to study, e.g., spatial, temporal and spatiotemporal solitons \cite{stegeman:1996,buryak:2002,Malomed:2005}, high-energy pulse compression \cite{liu:1999,ashihara:2002,moses:2006,zhou:2011}, supercontinuum generation \cite{Langrock:2007}, and all-optical signal processing \cite{Langrock:2006}.

Recently we investigated high-energy pulse compression through few-cycle solitons generated in cascaded SHG, and showed that the soliton can couple to dispersive waves \cite{bache:2008,bache:2010e}. 
Dispersive waves were first predicted as a consequence of perturbing a stable temporal soliton in the nonlinear Schr{\"o}dinger equation (NLSE) \cite{Wai:1986}. The first experimental observations came shortly after in a mode-locked dye laser \cite{Wise:1988} and in single-mode fibers \cite{beaud:1987,Gouveia-Neto:1988}; both systems are well-described by the NLSE. The dispersive wave is linear in nature and Ref. \cite{akhmediev:1995} showed that its spectral location is a result of a phase-matching condition to the soliton, and that its strength relates to the overlap to the soliton spectrum. There it was also pointed out that the dispersive wave shed by the soliton is reminiscent of optical Cherenkov radiation, and it has recently found applications in supercontinuum generation \cite{husakou:2001}, for ultra-short pulse synthesis \cite{krauss:2010}, broadband frequency combs \cite{Chang:2010} and UV femtosecond pulse generation \cite{im:2010,joly:2011}.

Cascaded SHG can induce both self- and cross-phase modulation cubic nonlinear terms \cite{clausen:1997,ditrapani:2001,corney:2001}, and in the limit of a weak SH the FW is well-described by an NLSE, just like the cases above. In Ref. \cite{bache:2010e} we used this property to show that in self-defocusing cascaded SHG the Cherenkov wave is emitted at longer wavelengths than the soliton. Self-defocusing  solitons namely form in the normal group-velocity dispersion (GVD) regime \cite{ashihara:2002}, and the Cherenkov wave is emitted in the non-solitonic anomalous GVD regime at longer wavelengths.

Important C-H, N-H and O-H stretching modes reside around $\lambda=3\mic$, and probing or controlling these require access to energetic few-cycle mid-IR pulses. Unfortunately, the beta barium borate crystal studied in Ref. \cite{bache:2010e} has an absorption edge around $\lambda=3\mic$. Recently we showed experimentally that noncritical cascaded SHG in lithium niobate (LN) can support near-IR few-cycle solitons \cite{zhou:2011}, and the predicted phase-matching to mid-IR Cherenkov waves \cite{bache:2010e} was confirmed numerically. Here we give a detailed numerical investigation of these mid-IR Cherenkov waves: they lie in the $\lambda=2.2-4.5\mic$ range (the exact resonance point depends on the soliton wavelength and strength), they can be long-pass filtered, have excellent pulse quality, and are born with almost transform-limited few-cycle duration. The conversion efficiency is 1-25\%, depending on how far the Cherenkov wavelength is from the soliton wavelength, implying that this scheme could be used as an efficient source of energetic few-cycle mid-IR pulses.


\section{Theory}

In Ref. \cite{bache:2010e} we used that cascaded SHG with an effective self-defocusing nonlinearity under suitable conditions can be reduced to a normalized NLSE for the FW field $U_1$
\begin{equation}
    \label{eq:fh-shg-nlse-nonlocal2}
  \left[i\frac{\partial}{\partial\xi}+ \Dm_1'\right]U_1
  -\neffs^2U_1|U_1|^2 = 0
  , \quad \neffs^2=\ld\frac{\omega_1}{c}\Iin |n_{\rm eff}^I| =\ld\frac{\omega_1}{c}\Iin(|n_{\rm casc}^I|-n_{\rm Kerr,el}^I),
\end{equation}
The effective soliton order $\neffs$ \cite{bache:2007} appears due to the following normalization choices $\xi=z/\ld$, $\tau'=\tau/\tin$, where $\tin$ is the input pulse duration, $\ld=\tin^2/|\kpp_1|$ is the dispersion length, and $|U_1|^2$ is normalized to the input intensity $\Iin$.
%
The normalized dispersion operator is $ \Dm_1'=\sum_{m=2}^{\infty} i^m \delta_1^{(m)}\frac{\partial^m}{\partial \tau'^m}$, where $\delta_1^{(m)}\equiv \ld k_1^{(m)}(\tin^{m}m!)^{-1} $, and $k_1^{(m)}=d^mk_1(\omega)/d\omega^m|_{\omega=\omega_1}$. Interaction in a bulk medium implies $k_1(\omega)=n_1(\omega)\omega/c$, where $n_1(\omega)$ is the FW linear refractive index. The cascaded nonlinear response is
$n_{\rm casc}^I=-2 \omega_1 d_{\rm eff}^2/[c^2\varepsilon_0n_1^2n_2 \Delta k]$ \cite{bache:2010e}, and $n_{\rm Kerr,el}^I$ is the electronic Kerr nonlinear refractive index. Besides being in the cascading limit ($|\Delta k L|\gg 1$) with a self-defocusing cascaded nonlinearity ($\Delta k>0$), the conditions for using the above model to describe the cascaded interaction are \cite{bache:2010e} (a) a broad-band cascaded nonlinear response, which requires $\Delta k>d_{12}^2/2\kpp_2$ (the so-called stationary regime) \cite{bache:2007a}, (b) a modest effective soliton order, which serves to reduce self-steepening effects (either direct or through cascading \cite{moses:2006b}), (c) a weak SH, in order to minimize cascaded \cite{clausen:1997} and Kerr \cite{bache:2007} cross-phase modulation terms, and (d) a negligible Raman response (for the LN crystal we consider in this paper we calculated the characteristic Raman time $T_R=0.4$ fs, which justifies this assumption).

We seek an overall effective self-defocusing nonlinearity $n_{\rm eff}^{I}<0$ so solitons can be excited in the normal GVD regime. In the simplest case a soliton will shed Cherenkov radiation according to the phase-matching condition $k_{\rm dw}(\omega_{\rm dw})-k_{\rm sol}(\omega_{\rm dw})=0$. In cascaded SHG the FW forms a soliton at some frequency $\omega_{\rm sol}$. The dispersion relation reflects its nondispersive nature: $k_{\rm sol}(\omega)=k_1(\omega_{\rm sol})+(\omega-\omega_{\rm sol})/v_{g,{\rm sol}}+q_{\rm sol}$, where $v_{g,\rm sol}=1/k_1^{(1)}(\omega_{\rm sol})$ is the soliton group velocity and $q_{\rm sol}$ is the soliton wave number. The Cherenkov wave dispersion is simply determined by the FW wavenumber $k_{\rm dw}(\omega)=k_{1}(\omega)$.  This gives the phase matching condition \cite{husakou:2001}
$  k_1(\omega_{\rm dw})- k_1(\omega_{\rm sol}) -(\omega_{\rm dw}-\omega_{\rm
  sol})k_1^{(1)}(\omega_{\rm sol})- q_{\rm sol}=0$.
The soliton wavenumber can be estimated from Eq. (\ref{eq:fh-shg-nlse-nonlocal2}) as $q_{\rm sol}=n_{\rm eff}^II_{\rm sol}\omega_{\rm sol}/2c$ \cite{bache:2010e}. Predicting accurately the soliton intensity $I_{\rm sol}$ is difficult, but for low soliton orders its contribution is minimal (see Ref. \cite{bache:2010e}). Therefore the phase-matching condition can be approximated by \cite{bache:2008}
\begin{equation}\label{eq:pm}
k_1(\omega_{\rm dw})- k_1(\omega_{\rm sol}) -(\omega_{\rm dw}-\omega_{\rm
  sol})k_1^{(1)}(\omega_{\rm sol})=0
\end{equation}

%

In Fig. \ref{fig:filter}(a) the phase-matching condition Eq. (\ref{eq:pm}) is shown for 5\% MgO doped LN (MgO:LN) X-cut ($\theta=\pi/2$) for noncritical (type 0, $ee\rightarrow e$) interaction. Pump wavelengths of $\lambda_1=1.2-1.8\mic$ should generate Cherenkov waves in the desired region around $\lambda=3\mic$. A prerequisite is exciting a soliton broadband enough to spectrally overlap the Cherenkov resonant wavelength, and in Ref. \cite{zhou:2011} we demonstrated this experimentally: a few-cycle self-defocusing soliton was formed in LN through noncritical cascaded SHG pumped with 50 fs pulses at $\lambda_1=1.3\mic$. The key factor is a large and ultrafast self-defocusing cascaded nonlinearity and we showed that it remains so at other pump wavelengths as well. Consider $\lambda_1=1.65\mic$ (the case shown below): despite a huge phase mismatch, $\Delta k=k_2-2k_1=283~\rm mm^{-1}$ corresponding to $11\mic$ coherence length, the strong effective nonlinearity $d_{\rm eff}\simeq 20~\rm pm/V$ \cite{Shoji:1997} is able to create a large negative cascaded nonlinearity $n_{\rm SHG}^I\simeq  -40\times 10^{-20}~\rm m^2/W$ without using quasi-phase matching. A large group-velocity mismatch (GVM) $d_{12}=-250~\rm fs/mm$ at $\lambda_1=1.65\mic$ gives a just $120\mic$ walk-off length for a 50 fs pulse, which should render femtosecond cascading inefficient, but the large phase mismatch ensures an ultrafast cascaded nonlinearity: it is of nonresonant nature (practically instantaneous with a sub-fs response time) as the interaction occurs in the so-called stationary regime \cite{bache:2007a} $\Delta k>\Delta k_{\rm sr}=d_{12}^2/2\kpp_2=90~{\rm mm^{-1}}$.

\section{Numerical simulations}

The plane-wave numerical simulations below are based on the slowly evolving wave approximation \cite{moses:2006b,bache:2007}. Since there are no direct measurements in the literature of the electronic and Raman Kerr nonlinear strengths of LN (see discussion in Refs. \cite{bache:2010,zhou:2011}), we use $n_{\rm Kerr}^I=30\times 10^{-20}~\rm m^2/W$ at $\lambda=1.30\mic$, and a Raman fraction $f_R=52\%$ \cite{zhou:2011}. Thus,  $n_{\rm Kerr,el}^I=(1-f_R) n_{\rm Kerr}^I=14.4\times 10^{-20}~\rm m^2/W$, making the effective nonlinearity self-defocusing for $\lambda_1>0.8\mic$ \cite{zhou:2011} and solitons can form in the normal GVD regime ($\lambda<\lambda_{\rm ZD}=1.92\mic$).

\begin{figure}[tb]
\begin{center}
\includegraphics[height=3.4cm]{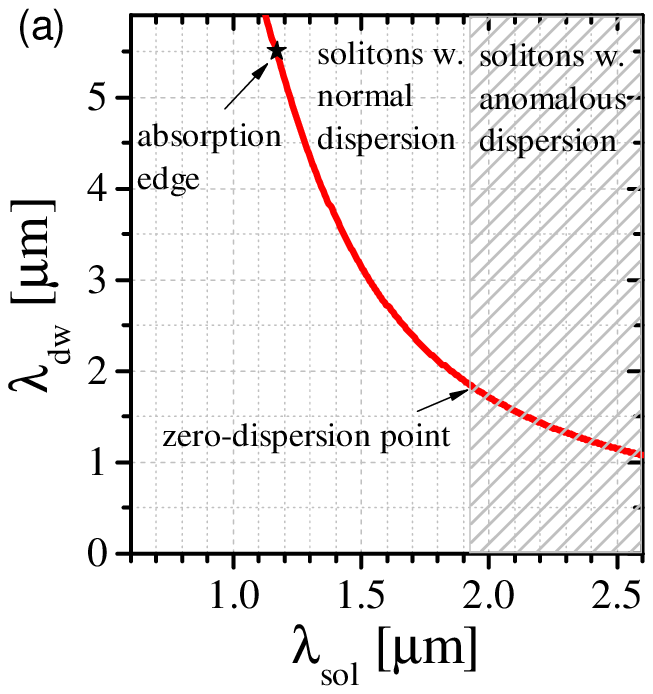}
\includegraphics[height=3.4cm]{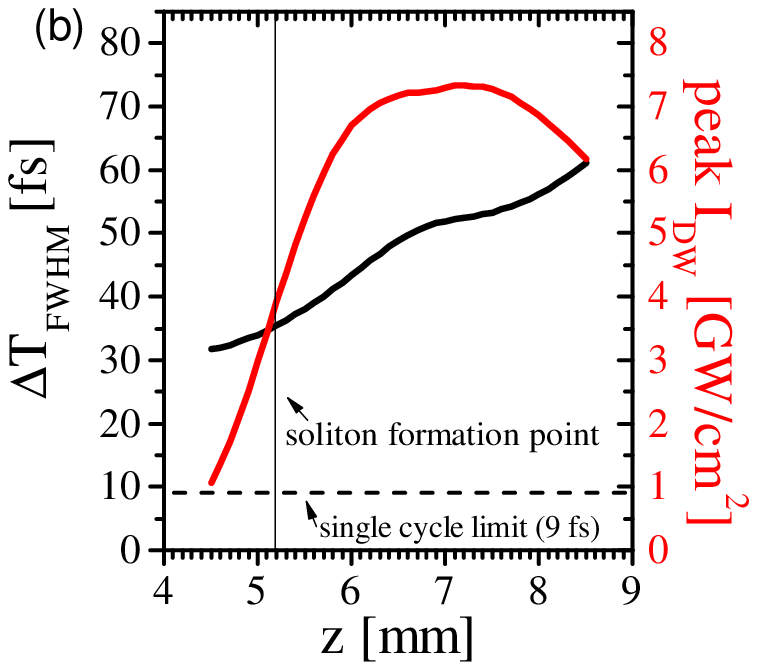}
\includegraphics[height=3.4cm]{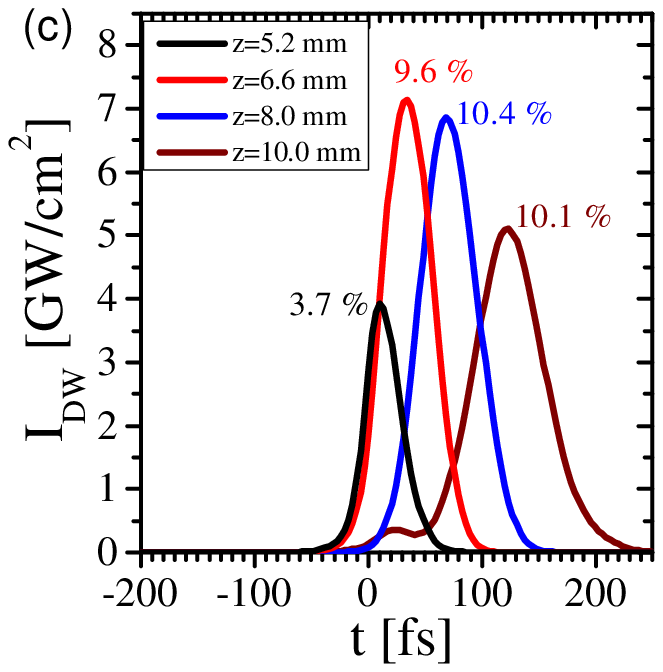}
\includegraphics[height=4.3cm]{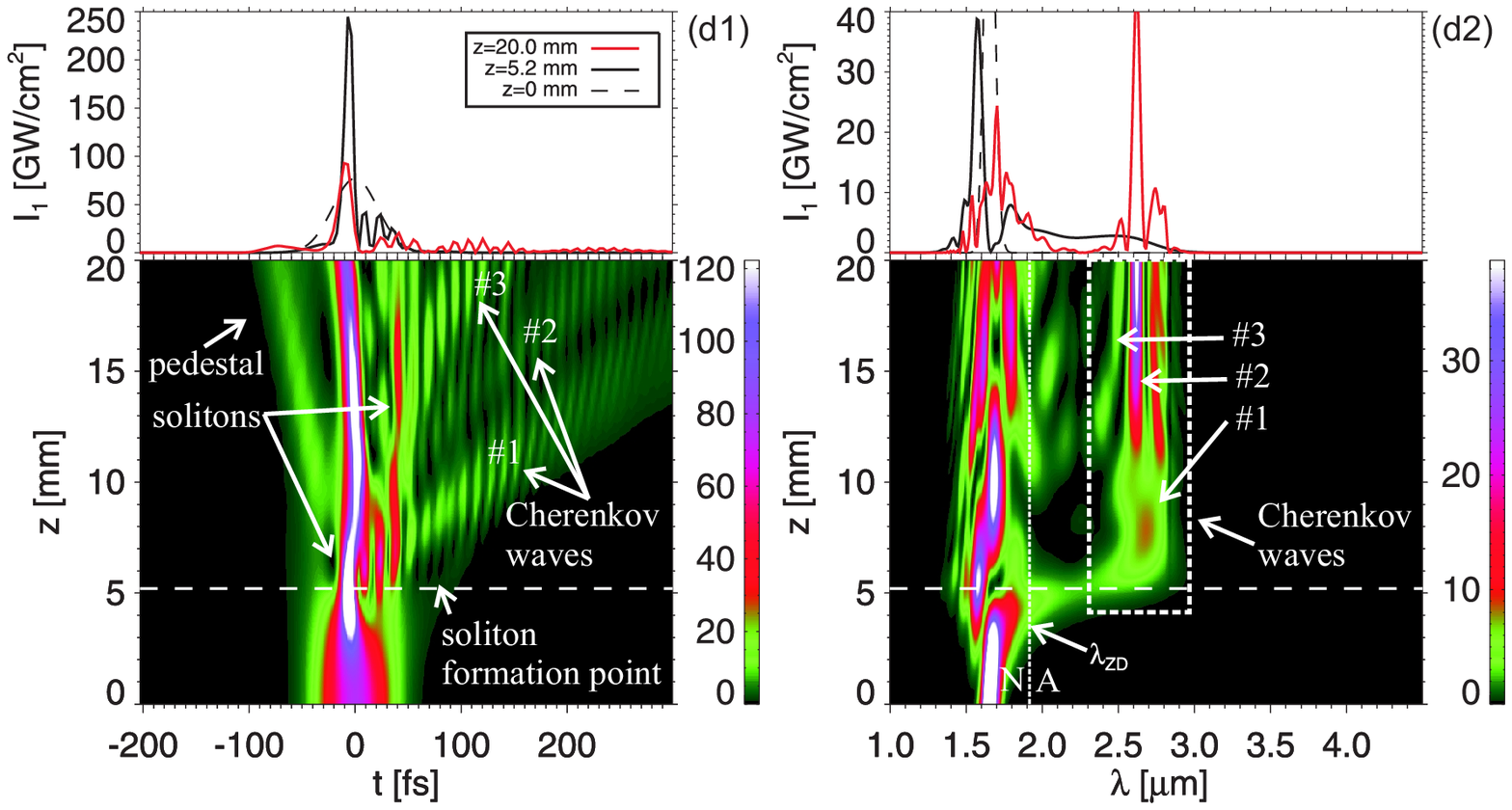}
\includegraphics[height=4.3cm]{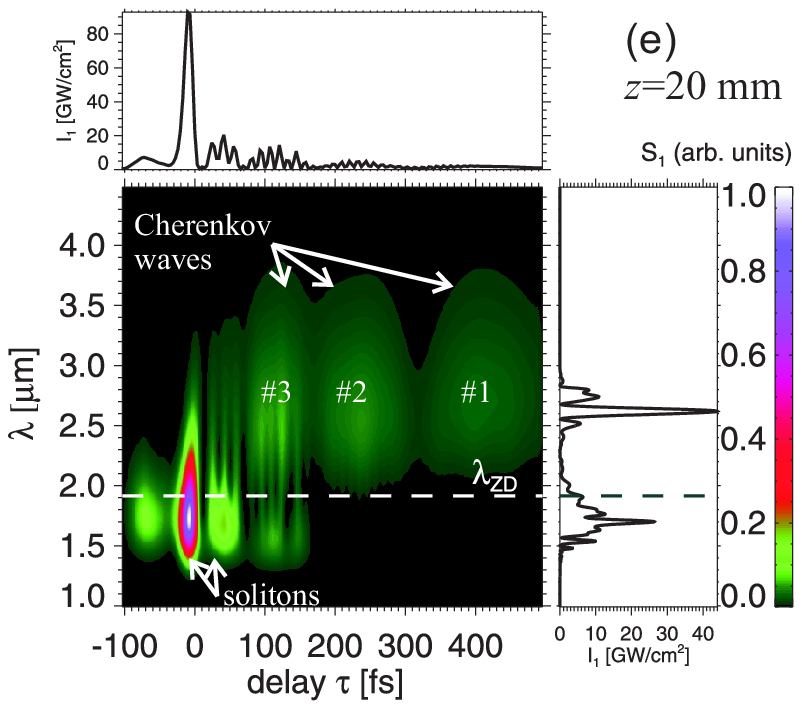}
\vspace{-16pt}
\end{center}
\caption{\label{fig:filter}
(a) Cherenkov phase-matching condition vs. soliton wavelength, Eq. (\ref{eq:pm}), for cascaded noncritical SHG in MgO:LN (using the Sellmeier equation at 24.5$^\circ$C from Ref. \cite{gayer:2008}). Data from Ref. \cite{bache:2010e}.
(b-d) Simulation of noncritical cascaded SHG in a 20 mm MgO:LN crystal using $\lambda_1=1.65\mic$, 50 fs FWHM, $\neffs=3.0$, $\Iin=77~{\rm GW/cm^2}$. (b) Intensity and pulse duration after longpass filtering the FW ($\lambda>2.5\mic$ transmitted). (c) The filtered Cherenkov wave at various propagation stages (the conversion efficiency is indicated above each stage). (d) shows the FW time trace and spectrum. In the spectrum 'A' and 'N' mark anomalous and normal dispersion regimes, and the color scale is saturated for clarity. (e) XFROG spectrogram (linear scale) at $z=20$ mm calculated with a 10 fs gating pulse.}
\end{figure}

Figure \ref{fig:filter} (b-d) shows a numerical simulation using $\lambda_1=1.65\mic$ 50 fs FWHM Gaussian input pulses with $\neffs=3.0$: (d1) shows that after 5.2 mm a compressed few-cycle soliton forms (10 fs FWHM, sub-2 cycles), after which a Cherenkov wave (marked \#1) is formed. It soon detaches from the soliton as its group velocity is much slower. In the spectrum (d2) this Cherenkov wave has a peak at $\lambda_{\rm dw}=2.65\mic$. In Fig. \ref{fig:filter}(b) we use a longpass edge filter at $2.5\mic$ to isolate the Cherenkov wave: it is less than four optical cycles (35 fs FWHM) when formed, it is broadband (245 nm FWHM, or 350 $\rm cm^{-1}$) and is near the transform limit. It then spreads out temporally but importantly the maximum peak intensity grows because the soliton keeps feeding radiation into the Cherenkov wave. This causes a GVM-induced broadening of the Cherenkov wave, which is in addition to that caused by pure GVD. The soliton then relaxes (decompresses) so at $z=8$ mm it is too weak to feed radiation into the Cherenkov wave. At this point the Cherenkov wave detaches temporally from the soliton, and the observed drop in intensity is explained by dispersive broadening. Note that unlike the self-focusing case, where Raman red-shifting causes the soliton to slow down and continuously collide with the Cherenkov wave \cite{gorbach:2007b}, here the Raman red-shift actually speeds up the soliton and the Cherenkov wave detaches for good without any further interaction \cite{bache:2010e}. Figure \ref{fig:filter}(c) shows the filtered Cherenkov wave at different stages: it has and excellent near-Gaussian pulse quality, and the conversion efficiency starts at 4\% and increases to over 10\% with further propagation. Eventually soliton fission occurs \cite{kodama:1987} creating a delayed minor soliton, which at $z=10-12$ mm emits a new Cherenkov wave (marked \#2) with a similar bandwidth and center wavelength as \#1. This causes the mid-IR spectrum to split in two. The Raman-induced frequency red-shift of the major soliton is larger than that of the minor soliton because its duration is shorter, so its Cherenkov resonance point should be closer to $\lambda_{\rm ZD}$ than that of the minor soliton. This is not the case, and we believe that it is because the spectral slope of the minor soliton is much steeper, pushing the Cherenkov wave to shorter wavelengths. At $z=15$ mm the main soliton has recompressed and it emits a new burst of Cherenkov radiation, marked \#3. An XFROG spectrogram \cite{bache:2010e} at $z=20$ mm is shown in (e), elucidating the temporal and spectral location of the interacting waves.

\begin{figure}[tb]
\begin{center}
\includegraphics[height=3.4cm]{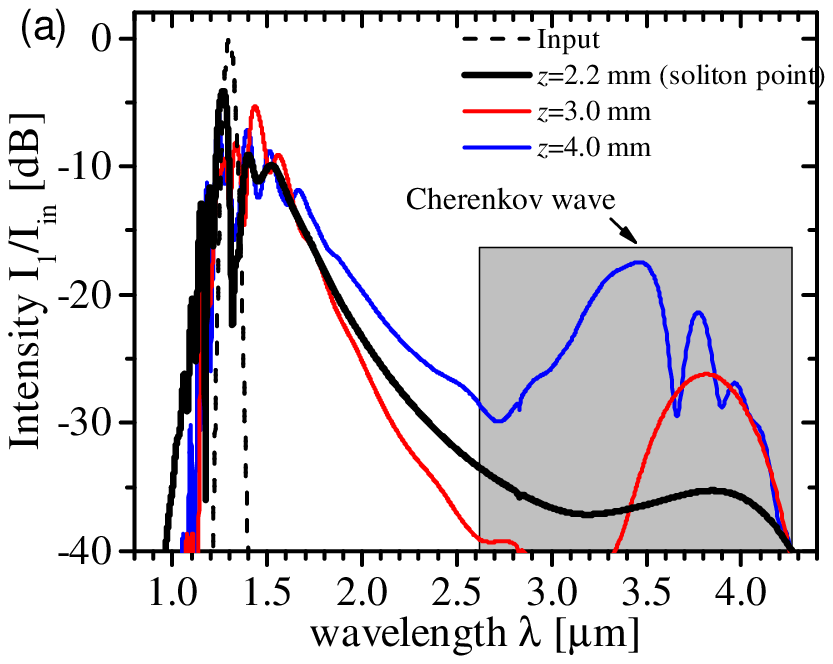}
\includegraphics[height=3.4cm]{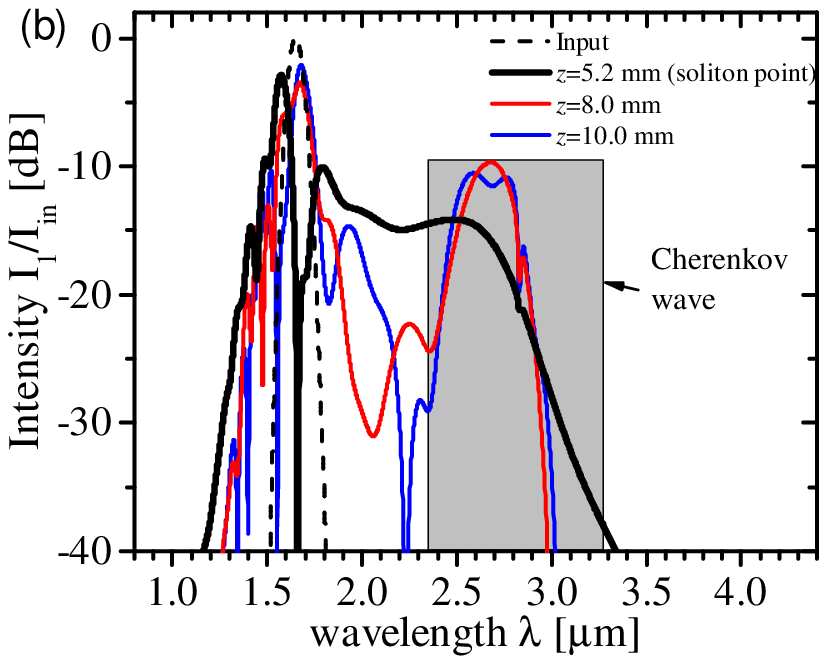}
\includegraphics[height=3.4cm]{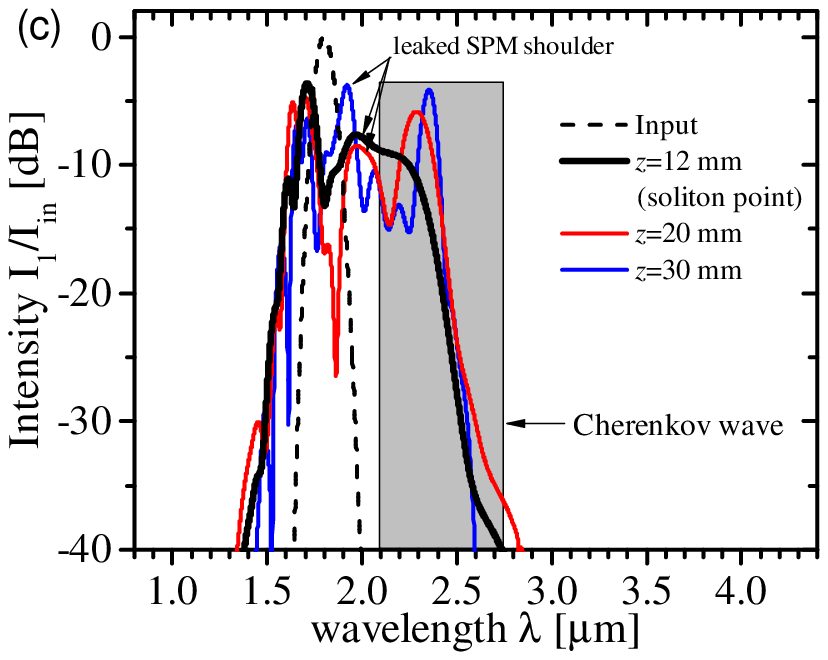}
\vspace{-16pt}
\end{center}
\caption{\label{fig:wavelength} Spectral contents at the soliton formation point and at later stages on a log-scale for (a) $\lambda_1=1.3\mic$, (b) $\lambda_1=1.65\mic$ and (c) $\lambda_1=1.8\mic$. In all three cases the initial pulse has 50 fs FWHM and $\neffs=3.0$ was kept constant implying an input intensity of (a) $\Iin=215~{\rm GW/cm^2}$, (b) $\Iin=77~{\rm GW/cm^2}$, and (c) $\Iin=35~{\rm GW/cm^2}$.}
\end{figure}

In Fig. \ref{fig:wavelength} the spectral evolution from Fig. \ref{fig:filter} is compared to other pump wavelengths. In all cases the peak Cherenkov intensity grows upon propagation, and later several peaks appear because new Cherenkov waves are emitted by soliton recompression and fission. When pumping far from $\lambda_{\rm ZD}$, the Cherenkov radiation is very weak, see Fig. \ref{fig:wavelength}(a) where $\lambda_1=1.3\mic$, but in return it has 2-cycle duration. As the pump wavelength is increased, the Cherenkov wave shifts to lower wavelengths since the resonance wavelength changes, and its bandwidth reduces. As less bandwidth is required to sustain shorter-wavelength few-cycle pulses, a 4-cycle near-IR Cherenkov wave with $\lambda_{\rm dw}=2.1\mic$ can still be isolated in case (c) where $\lambda_1=1.8\mic$. The Cherenkov wave also becomes more intense when the pump wavelength approaches $\lambda_{\rm ZD}$ because the spectral overlap with the soliton is larger \cite{akhmediev:1995}. In (c) the pump is quite close to $\lambda_{\rm ZD}$, and the red-shifted SPM shoulder leaks into the anomalous dispersion region (see also Ref. \cite{bache:2010e}). This complicates the propagation dynamics after the initial soliton formation in a similar way as soliton fission. Despite this, the efficiency is very high, up to 25\%. Finally, note that a longer interaction length is needed close to $\lambda_{\rm ZD}$ because the dispersion length increases.

\section{Conclusion}

Trough numerical simulations of ultrafast noncritical cascaded SHG in LN we showed that few-cycle solitons can be formed that shed near- to mid-IR optical Cherenkov radiation in the $\lambda=2.2-4.5\mic$ range with few-cycle duration, excellent pulse quality, and a high conversion efficiency (up to 25\% was observed). We recommend a low soliton order to keep perturbations to the main soliton at a minimum so it can boost the Cherenkov wave over an extended propagation stage without distorting it. The Cherenkov waves could be isolated using a longpass filter, and 2-cycle mid-IR pulses were found when pumping far from the zero-dispersion wavelength $\lambda_{\rm ZD}$. This case has a low conversion efficiency, but it increases when pumping close to $\lambda_{\rm ZD}$ where the Cherenkov wave shifts to shorter wavelengths, and the pulse duration increase slightly to 3-4 cycles when formed. Alternative methods for generating energetic few-cycle mid-IR pulses -- like optical rectification \cite{Likforman:2001} or noncollinear optical parametric amplification \cite{Brida:2008} -- are either inefficient or very complex. Thus, cascaded SHG might provide an efficient bridge between near-IR femtosecond laser technology and ultrashort energetic mid-IR pulses.

\end{document}